# Chemical-potential-based lattice Boltzmann method for nonideal fluids

Binghai Wen[1,2], Xuan Zhou[1], Bing He[1], Chaoying Zhang[1,*], Haiping Fang[2]

[1]Guangxi Key Lab of Multi-source Information Mining & Security, Guangxi Normal University, Guilin 541004, China

[2]Division of Interfacial Water and Key Laboratory of Interfacial Physics and Technology, Shanghai Institute of Applied Physics, Chinese Academy of Sciences, Shanghai 201800, China

The chemical potential is an effective way to drive a phase transition or express wettability. We present a chemical-potential-based lattice Boltzmann model to simulate multiphase flows. The nonideal force is directly evaluated by a chemical potential. The model theoretically satisfies thermodynamics and Galilean invariance. The computational efficiency is improved because the calculation of the pressure tensor is avoided. We have derived several chemical potentials of the popular equations of state from the free-energy density function. An effective chemical-potential boundary condition is implemented to investigate the wettability of a solid surface. Remarkably, the numerical results show that the contact angle can be linearly tuned by the surface chemical potential.

*Corresponding author. E-mail: zhangcy@gxnu.edu.cn

# I. INTRODUCTION

Numerical simulation of multiphase flow is one of the most successful applications of the lattice Boltzmann method (LBM) [1-4]. Originating from the cellular automaton concept and kinetic theory, the intrinsic mesoscopic properties make LBM outstanding in modeling complex fluid systems involving interfacial dynamics [5-7] and phase transitions [3,4]. Shan and Chen proposed the widely-used pseudopotential LBM, in which the intermolecular force is mimicked by a density-dependent interparticle potential [8]. However, theoretical analyses showed that its mechanical stability solution agrees with thermodynamics only if the effective mass takes the strict functional form $\psi(\boldsymbol{x}) = \psi_0 \exp(-\rho/\rho_0)$, where $\rho$ is the density, $\rho_0$ and $\psi_0$ are the reference density and effective mass, respectively [8,9]. Therefore, the model is indeed thermodynamically inconsistent, regardless of whether the most commonly used functional form $\psi(\boldsymbol{x}) = 1 - \exp(-\rho/\rho_0)$ or subsequent improvement $\psi(\boldsymbol{x}) = \sqrt{2(p - \rho c_s^2)/G}$ is employed, here $p$ is the pressure given by an equation of state, $G$ a factor to ensure that the whole term inside the square root is positive [8-12]. This inconsistency was also clearly presented by the deviations between the numerical two-phase coexistence densities and the predictions from the Maxwell equal-area construction [10,12,13]. Recently, Li *et al.* [12] proposed a forcing term scheme and Khajepor *et al.* [14] proposed a multipseudopotential interaction to achieve thermodynamic consistency. However, a fitting procedure is necessary in both schemes to adapt to a specific equation of state (EOS) [3]. The Shan-Chen model was also improved for simulations with large density ratio and high stability [15,16]. Swift *et al.* proposed another popular multiphase model, which is based on free energy and incorporates a nonideal thermodynamic pressure tensor into the second equilibrium moment [17]. Inamuro *et al.* used an asymptotic analysis to accommodate the lack of Galilean invariance and extended the model to simulate multiphase flows with large density ratio [18,19]. Zheng *et al.* correctly recovered the Cahn–Hilliard equation to capture the phase interface and Shao *et al.* further considered the effect of local density variation in the momentum equation [20,21]. Wang *et al.* proposed a multiphase lattice Boltzmann flux solver for incompressible multiphase flows with large density ratio and high Reynolds number [22,23]. Typically, these models use two kinds of distribution functions, one evolves the pressure and velocity field by the standard lattice Boltzmann equation and the other captures the phase interface by the Cahn-Hilliard equation. Recently, Wen *et al.* directly computed the nonideal force from the free energy and proposed a multiphase flow model, which satisfies thermodynamics and Galilean invariance [13]. A similar nonideal force evaluation was also used in the entropic lattice Boltzmann method [24].

Chemical potential (CP) is the partial molar Gibbs free energy at constant pressure [25]. Some studies of multiphase flows used the chemical potential to present the phase equilibrium condition and obtain the density profile along the interfacial normal direction when the bulk free energy density took the simplest double-well form [20,25,26]. However, both the Onsager and Stefan-Maxwell formulations of irreversible thermodynamics recognize that the chemical potential gradient is the driving force for isothermal mass transport [27,28]. Movement of molecules from higher chemical potential to lower chemical potential is accompanied by a release of free energy, and at the minimum free energy, chemical equilibrium or phase equilibrium is achieved. Therefore, the chemical potential is an effective way to drive a phase transition or express surface wettability.

In this paper, we present a multiphase lattice Boltzmann model by introducing a chemical potential to directly evaluate the nonideal force. The computational efficiency is improved as the calculation of the pressure tensor is avoided. We implement a chemical-potential boundary condition to effectively investigate the surface wettability and find that the contact angle is almost linear-tunable by the chemical potential of the solid surface.

## II. THEORY AND MODEL

Considering a nonideal fluid system, the free-energy functional within a gradient-squared approximation is [13,17,29]

$$\Psi = \int (\psi(\rho) + \frac{\kappa}{2}|\nabla\rho|^2) d\bm{x}, \quad (1)$$

where the first term is the bulk free-energy density at a given temperature with the density $\rho$ and the second term gives the free-energy contribution from density gradients in a inhomogeneous system. The free energy function in turn determines the diagonal term of the pressure tensor

$$p(\bm{x}) = p_0 - \kappa\rho\nabla^2\rho - \frac{\kappa}{2}|\nabla\rho|^2, \quad (2)$$

with the general expression of equation of state

$$p_0 = \rho\psi'(\rho) - \psi(\rho). \quad (3)$$

The full pressure tensor can be written as

$$P_{\alpha\beta}(\bm{x}) = p(\bm{x})\delta_{\alpha\beta} + \kappa\frac{\partial\rho}{\partial x_\alpha}\frac{\partial\rho}{\partial x_\beta}. \quad (4)$$

where $\delta_{\alpha\beta}$ is the Kronecker delta function. The excess pressure, namely the nonideal force, with respect to the ideal gas expression can be directly computed [9,13]

$$\bm{F}(\bm{x}) = -\nabla\cdot\vec{\bm{P}}(\bm{x}) + \nabla\cdot\vec{\bm{P}}_0(\bm{x}), \quad (5)$$

where $\vec{\bm{P}}_0 = c_s^2\rho\vec{\bm{I}}$ is the ideal-gas EOS. The nonideal force evaluation is thermodynamically consistent and Galilean invariant [13].

The above multiphase model is directly derived from thermodynamics; however, the evaluation of the pressure tensor and its divergence has high complexity, both temporally and spatially. Moreover, the pressure tensor cannot be conveniently used to express the wettability of a solid surface. Here, we introduce a chemical potential to evaluate the nonideal force. For a van der Waals (VDW) fluid, the chemical potential can be derived from the free-energy density function [20,25,29]

$$\mu = \frac{\partial\Phi}{\partial\rho} - \nabla\cdot\frac{\partial\Phi}{\partial(\nabla\rho)} \quad (6)$$

where

$$\Phi(\rho) = \psi(\rho) + \frac{\kappa}{2}(\nabla\rho)^2.\tag{7}$$

Thus, the chemical potential is computed by the density and free-energy density

$$\mu = \psi'(\rho) - \kappa\nabla^2\rho \tag{8}$$

From Eqs. (2) and (4), we find

$$P_{\alpha\beta} = [p_0 - \kappa\rho\nabla^2\rho - \frac{\kappa}{2}(\nabla\rho)^2]\delta_{\alpha\beta} + \kappa\frac{\partial\rho}{\partial x_\alpha}\frac{\partial\rho}{\partial x_\beta}.\tag{9}$$

Substitution of Eq. (3) into Eq. (9), the partial derivative of the pressure tensor is written as

$$\frac{\partial}{\partial x_\beta}P_{\alpha\beta} = \frac{\partial}{\partial x_\alpha}[\rho(\psi'(\rho)-\kappa\nabla^2\rho)-\psi(\rho)] - \frac{\partial}{\partial x_\alpha}[\frac{\kappa}{2}(\nabla\rho)^2] + \kappa\frac{\partial}{\partial x_\beta}(\frac{\partial\rho}{\partial x_\alpha}\frac{\partial\rho}{\partial x_\beta})\tag{10}$$

Taking Eq. (8) and after some simple manipulations, a simple relationship between the divergence of the pressure tensor and the gradient of the chemical potential can be obtained

$$\nabla\cdot\vec{P} = \rho\nabla\mu.\tag{11}$$

Substituting Eq. (11) into Eq. (5), the nonideal force can be evaluated by the chemical potential in the form

$$\boldsymbol{F}(\boldsymbol{x}) = -\rho\nabla\mu + \nabla\cdot\vec{P}_0(\boldsymbol{x}).\tag{12}$$

Then, the nonideal force is incorporated into the lattice Boltzmann equation (LBE), which is fully discretized in space, time and velocity. Several collision operators distinguish the variants of the LBE, such as the single-relaxation-time model [30-32], the multiply-relaxation-time model [33], the two-relaxation-time model [34], and the entropic lattice Boltzmann equation [35]. The single-relaxation-time version can be concisely written as

$$f_i(\boldsymbol{x}+\boldsymbol{e}_i,t+1) - f_i(\boldsymbol{x},t) = -\frac{1}{\tau}[f_i(\boldsymbol{x},t) - f_i^{(eq)}(\boldsymbol{x},t)],\tag{13}$$

where $f_i(\boldsymbol{x},t)$ is the particle distribution function at lattice site $\boldsymbol{x}$ and time $t$, $\boldsymbol{e}_i$ with $i = 0, ..., N$ is the discrete speed, $\tau$ the relaxation time, and $f_i^{(eq)}$ the

equilibrium distribution function

$$f_i^{(eq)}(\boldsymbol{x},t) = \rho\omega_i[1 + 3(\boldsymbol{e}_i \cdot \boldsymbol{u}) + \frac{9}{2}(\boldsymbol{e}_i \cdot \boldsymbol{u})^2 - \frac{3}{2}u^2], \tag{14}$$

where $\omega_i$ is the weighting coefficient and $\boldsymbol{u}$ the fluid velocity. The nonideal force acts on the collision process by a simple increase in the particle momentum in $f_i^{(eq)}$ in terms of the momentum theorem. The velocity in Eq. (14) is replaced by the equilibrium velocity $\boldsymbol{u}^{eq} = \boldsymbol{u} + \tau\boldsymbol{F}/\rho$ [36]. Correspondingly, the macroscopic fluid velocity is redefined by the averaged momentum before and after the collision $\boldsymbol{v} = \boldsymbol{u} + \boldsymbol{F}/2\rho$.

In the above deductions, it is clear that no specific EOS is presupposed and Eq. (12) is established only if an EOS follows the form of Eq. (3) and the corresponding chemical potential meets the definition of Eq. (6). The model is then versatile to simulate various multiphase flows by adapting to its EOS, such as that of van der Waals, Redlich-Kwong, Peng-Robinson, Carnahan-Starling. Eq. (12) is mathematically equivalent to the previous version, namely Eq. (5), based on the pressure tensor. Therefore the nonideal force evaluation based on the chemical potential also theoretically satisfies thermodynamics and Galilean invariance. Furthermore, the chemical-potential-based model is efficient in both computing time and memory, because it avoids computing the pressure tensor and its divergence, which are computationally heavy and require more memory in the 3D simulations of multiphase flows.

### III. SPECIFIC CHEMICAL POTENTIALS

For a given equation of state $p_0$, solving Eq. (3), which is a typical one-order linear ordinary differential equation, determinates the general solution of the free-energy density

$$\psi = \rho(\int \frac{p_0}{\rho^2} d\rho + C), \tag{15}$$

where $C$ is a constant, but it does not appear in the evaluation of the nonideal force below. We select several widely used EOSs and solve their free-energy densities and chemical potentials.

The VDW EOS is the most famous cubic EOS

$$p_0 = \frac{\rho RT}{1-b\rho} - a\rho^2, \tag{16}$$

where $R$ is the universal gas constant, $a$ the attraction parameter and $b$ the volume correction. The free-energy density and chemical potential are obtained by solving Eqs. (8) and (15)

$$\psi = \rho RT \ln(\frac{\rho}{1-b\rho}) - a\rho^2 \tag{17}$$

and

$$\mu = RT[\ln(\frac{\rho}{1-b\rho}) + \frac{1}{1-b\rho}] - 2a\rho - \kappa\nabla^2\rho. \tag{18}$$

The Redlich-Kwong (RK) EOS is generally more accurate than the VDW EOS by improving the attraction term

$$p_0 = \frac{\rho RT}{1-b\rho} - \frac{a\alpha(T)\rho^2}{(1+b\rho)}, \tag{19}$$

where $\alpha(T) = 1/\sqrt{T}$. The Soave modification (RKS) involves a more complicated temperature function, $\alpha(T) = [1+(0.480+1.574\omega-0.176\omega^2)(1-\sqrt{T_r})]^2$, where $\omega$ is the acentric factor. Both equations share the same expression of free-energy density and chemical potential,

$$\psi^{RK} = RT\rho\ln\frac{\rho}{1-b\rho} - \frac{a\alpha(T)}{b}\rho\ln(1+b\rho), \tag{20}$$

and

$$\mu^{RK} = RT\ln\frac{\rho}{1-b\rho} + \frac{RT}{1-b\rho} - \frac{a\alpha(T)}{b}\ln(1+b\rho) - \frac{a\alpha(T)\rho}{1+b\rho} - \kappa\nabla^2\rho. \tag{21}$$

The Peng-Robinson (PR) EOS is often superior in predicting liquid densities,

$$p_0 = \frac{\rho RT}{1-b\rho} - \frac{a\alpha(T)\rho^2}{1+2b\rho-b^2\rho^2}, \tag{22}$$

where the temperature function is $\alpha(T) = [1+(0.37464+1.54226\omega-0.26992\omega^2)\times(1-\sqrt{T/T_c})]^2$. Its free-energy density and chemical potential are

$$\psi^{PR} = RT\rho \ln\frac{\rho}{1-b\rho} - \frac{a\alpha\rho}{2\sqrt{2}b}\ln\frac{\sqrt{2}-1+b\rho}{\sqrt{2}+1-b\rho}, \tag{23}$$

and

$$\mu^{PR} = RT\ln\frac{\rho}{1-b\rho} - \frac{a\alpha(T)}{2\sqrt{2}b}\ln\frac{\sqrt{2}-1+b\rho}{\sqrt{2}+1-b\rho} + \frac{RT}{1-b\rho} - \frac{a\alpha(T)\rho}{1+2b\rho-b^2\rho^2} - \kappa\nabla^2\rho. \tag{24}$$

The Carnahan-Starling (CS) EOS tends to give better approximations for the repulsive term

$$p_0 = \rho RT\frac{1+b\rho/4+(b\rho/4)^2-(b\rho/4)^3}{(1-b\rho/4)^3} - a\rho^2, \tag{25}$$

for which the free-energy density and chemical potential are

$$\psi^{CS} = RT\rho[\frac{3-b\rho/2}{(1-b\rho/4)^2} + \ln\rho] - a\rho^2, \tag{26}$$

and

$$\mu^{CS} = RT[\frac{3-b\rho/4}{(1-b\rho/4)^3} + \ln\rho + 1] - 2a\rho - \kappa\nabla^2\rho. \tag{27}$$

In the following simulations, the attraction parameter and the volume correction take $a=2/49$, $b=2/21$ for PR and RKS EOS and $a=1$, $b=4$ for CS EOS. The universal gas constant is $R=1$. The acentric factor $\omega$ is 0.344 for water and 0.011 for methane. In an effort to relate the numerical results to the real physical properties, we define the reduced variables $T_r = T/T_c$ and $\rho_r = \rho/\rho_c$, where $T_c$ is the critical temperature and $\rho_c$ the critical density.

The two-phase coexistence densities solved by the Maxwell equal-area construction are used as benchmarks to verify the thermodynamic consistency of the present multiphase model. Four EOSs, namely CS, RKS, PR for water and methane, are evaluated. From Fig. 1, the present numerical simulations are all in excellent agreement with the benchmarks. The liquid/gas density ratios computed by PR EOS for water and methane are more than 100. Only if a proper EOS is chosen, the present model is competent to accurately simulate real physical systems.

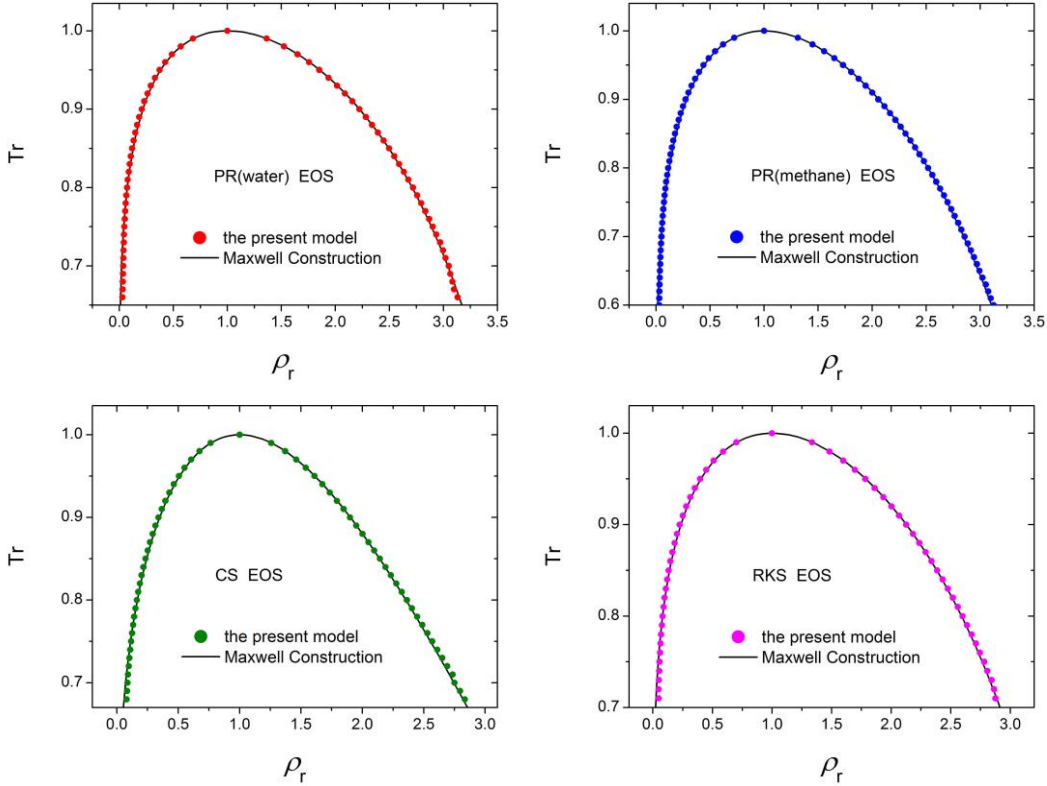

*Fig. 1. (Color online) Two-phase coexistence density curves of PR, CS and RKS EOSs compared with the theoretical predictions solved by the Maxwell equal-area construction.*

## IV. CHEMICAL-POTENTIAL BOUNDARY CONDITION

Chemical potential, which has a clear physical meaning, is a more effective way to express the wettability of a solid surface than density or effective mass (i.e., pseudopotential) [8,9,37,38]. An external chemical potential was once used at the surfaces of a confined system to study wetting [17]. In the present study, since the multiphase model is driven by a chemical potential, the implementation of the chemical-potential boundary condition is simple and quite natural. Endowing a solid surface with a specific chemical potential, the wettability, namely the interaction between the fluid and the solid, can be easily evaluated. Let us locate the straight interface of fluid and solid on a row of lattice nodes (y=1), which are treated as fluid nodes. The distribution functions on these interfacial fluid nodes still collide and stream. The bounce-back boundary condition is applied to mimic those distribution functions from the solid. The densities of the solid nodes (y=0) must be estimated to evaluate the nonideal force on the interfacial fluid nodes. They are only used in the calculation of the density gradient in Eq. (8). A simple weighted average scheme based on the neighbor fluid nodes is used here

$$\rho(x,0) = \frac{2}{3}\rho(x,1) + \frac{1}{6}\rho(x-1,1) + \frac{1}{6}\rho(x+1,1). \tag{28}$$

The wetting property of the solid surface is expressed by a chemical potential, which is assigned to the solid nodes (y=0) and acts on the gradient of chemical potential in Eq. (12).

## V. LINEAR-TUNABLE CONTACT ANGLE

Contact angle is an essential feature to reflect the wettability of a solid surface. In this section, we illustrate how the chemical-potential-based model can effectively simulate the wetting phenomena. The computational domain is a rectangular with the length $Dx = 500$ and the width $Dy = 200$. The temperature takes $T_r = 0.8$ and the drop radius is 50. The relaxation time is 1.3 and the value of $\kappa$ in Eq. (8) is 0.01. For the top side, the value of chemical potential is optional because the drop never touches it. Typically, the chemical potential on the top side takes the same value as that in its neighboring fluid node, which is always gas phase in the simulations. Without gravity, a droplet on a horizontal solid surface forms a perfect spherical cap. If the base and height of the droplet are L and H, the radius of the droplet is calculated by $R = (4H^2 + L^2)/8H$ and then the contact angle is obtained by the formula $\tan\theta = L/2(R-H)$.

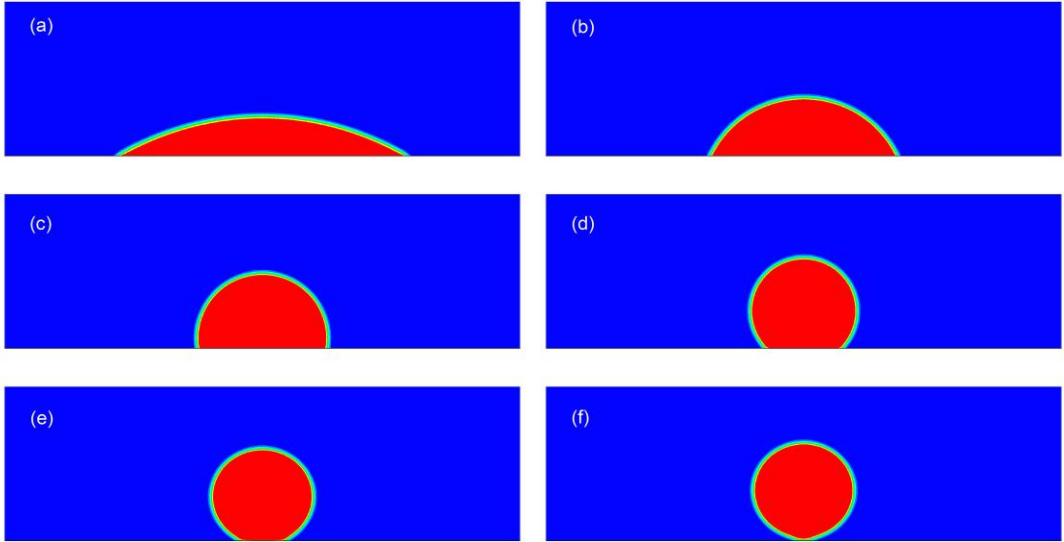

*Fig. 2. (Color online) Contact angles obtained on solid surfaces with different chemical potentials by using RKS EOS: (a) $\theta = 30^o$ at CP = -0.23, (b) $\theta = 60^o$ at CP = -0.13, (c) $\theta = 105^o$ at CP = 0, (d) $\theta = 135^o$ at CP = 0.09, (e) $\theta = 160^o$ at CP = 0.15, (f) $\theta = 180^o$ at CP = 0.18.*

   To illustrate the wettability of different surfaces, a sessile drop on a solid surface with specified chemical potentials is simulated using the RKS EOS. Fig. 2 demonstrates that the present model can effectively simulate a drop on the solid surfaces from hydrophilic to superhydrophobic.

   The relationship between the chemical potential and the contact angle is further investigated. The numerical simulations include four equations of state, namely RKS EOS, CS EOS, and PR EOSs for water and methane. As shown in Fig. 3, with the growth of the chemical potential of the solid surface, the simulating contact angles smoothly increase. Remarkably, the contact angle increases almost linearly with increasing chemical potential. Comparing with some multiphase models for which contact angle and controlled variables are related curvilinearly [9,39], a linear-tunable contact angle is very convenient to adjust the wettability of the solid surface.

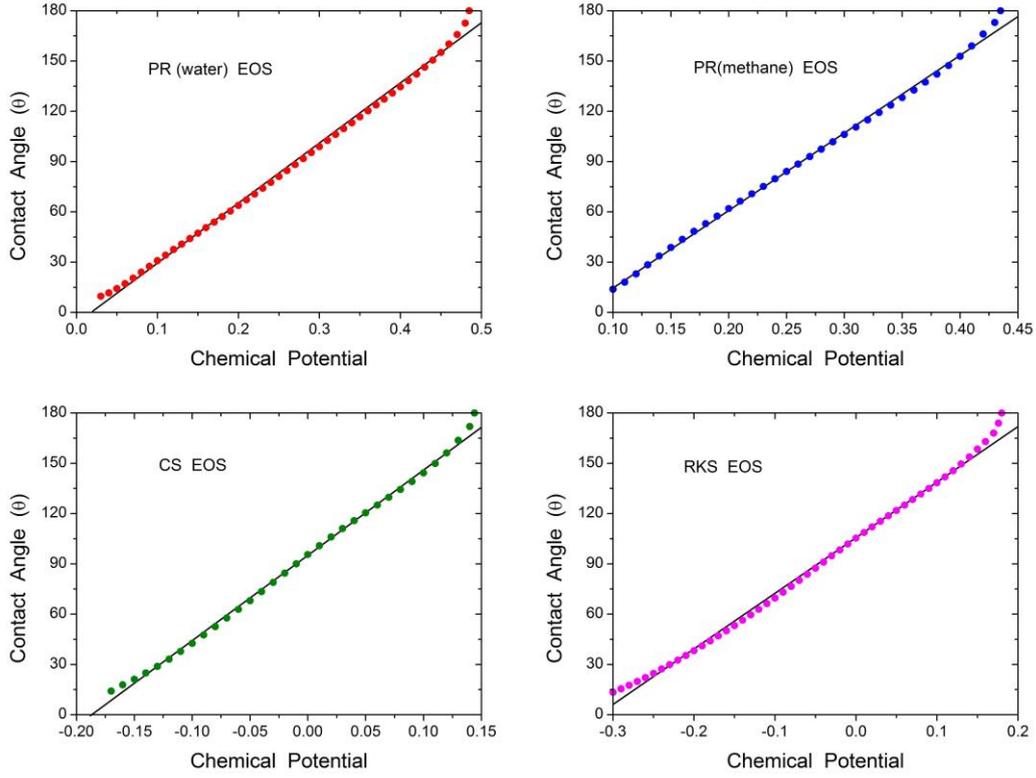

*Fig. 3. (Color online) Contact angle increasing with the chemical potential of the solid surface at the reduced temperature 0.8.*

## VI. CONCLUSION

We performed a multiphase lattice Boltzmann model for nonideal fluid based on chemical potential. The model is mathematically equivalent to the previous pressure-tensor-based multiphase model, which theoretically satisfies thermodynamics and Galilean invariance. A graceful relationship between the divergence of the pressure tensor and the gradient of the chemical potential has been obtained. Calculations of the pressure tensor and its divergence are avoided and thereby the computational efficiency is also improved. The model is versatile in simulating various multiphase flows by cooperating with different EOSs. Several chemical potentials of the popular equations of state have been derived through the bulk free-energy density function. To investigate the surface wettability, a chemical-potential boundary condition was implemented to effectively simulate the wetting solid surfaces from hydrophilic to superhydrophobic. Notably, the contact angle can be linearly tuned by the chemical potential of solid surface. The present method is expected to promote research of multiphase flows in real physical systems.

**ACKNOWLEDGMENTS**

This work was supported by the National Natural Science Foundation of China (Grant Nos. 11362003, 11462003, 11290164, 11162002), Guangxi Natural Science Foundation (Grant No. 2014GXNSFAA118018), Guangxi Science and Technology Foundation of College and University (Grant No. KY2015ZD017), Guangxi "Bagui Scholar" Teams for Innovation and Research Project, Guangxi Collaborative Innovation Center of Multi-source Information Integration and Intelligent Processing, CAS-Shanghai Science Research Center (Grant No. CAS-SSRC-YJ-2015-01).